\documentclass[11pt]{article}
\usepackage{amsmath,amssymb,amsfonts}
\usepackage[margin=1in]{geometry}
\usepackage{graphicx}    
\usepackage{bm}          
\usepackage{booktabs}
\usepackage{url}

\newcommand{\Rop}{\mathsf{R}}
\newcommand{\Top}{\mathsf{T}}
\newcommand{\Id}{\mathsf{I}}
\newcommand*\dif{\mathop{}\!\mathrm{d}}
\newcommand{\lstar}{\ell^{\ast}}
\newcommand{\Pinf}{P_{\infty}}
\newcommand{\PR}{P_{R}}

\newcommand{\Exp}{\mathsf{Exp}}

\newcommand{\avg}[1]{\left\langle #1 \right\rangle}

\begin{document}

\title{Finite slab first-passage statistics of Henyey--Greenstein scattering}
\author{Robert Cordery\thanks{Fairfield University, 246 Barn Hill Rd. Monroe CT; rcordery@fairfield.edu}
\and Claude Zeller\thanks{Claude Zeller Consulting LLC, Oceanside OR 97134; czeller@ieee.org}}
\date{June 30, 2026}
\maketitle

\begin{abstract}
A photon entering a plane-parallel scattering slab performs a random walk and eventually escapes through one of the two faces or is absorbed.
The standard model employs a Henyey-Greenstein phase function (HG) and an exponential step-length distribution \((\Exp)\).
Slab reflectance, transmittance, absorptance, and emergent angular distributions can be calculated in terms of random walk statistics. 
A central result is that the slab calculations factor into the order-resolved first-passage statistics of a half-space combined with the a factor for the slab thickness. 
Absorptance is derived from order-resolved walk statistics using the absorption rate. 
Two approaches are used.
 In the Monte Carlo (MC) approach, an extremely long random walk with many steps is efficiently generated without regard to any boundaries.
 The intersection of this walk with a large collection of target objects creates an ensemble of excursions of the objects.
The MC approach relies explicitly on the memoryless property of \(\Exp\) so that the  portion of the first and last steps inside the object follow the same length distribution as the walk steps. 
 The details of each excursion are recorded and any statistics can be extracted from the database of excursions.
 In particular, first passage statistics are extracted from this ensemble.
 In this work the objects are slabs with different positions and thicknesses.
 In the radiative-transfer (RT) approach the slab is divided into thin layers with scattering treated to first order in each layer.
 The RT equations are then directly integrated over the slab to give the desired first passage statistics, reflectance, transmittance, and absorptance.
%
%
The two methods agree to the Monte Carlo precision over the tested range of random walk parameters.

\end{abstract}



\section{Introduction}\label{sec:intro}
The reflection, transmission, and absorption of light by a finite scattering slab are classical problems in radiative transport, with applications ranging from coatings, paper, and biological tissue to snow, ice, and clouds.
For anisotropic scattering, the Henyey–Greenstein (HG) phase function~\cite{HenyeyGreenstein1941} provides a simple and widely used description of directional persistence through a single asymmetry parameter \(g\).
HG has been applied extensively to slab reflectance and photon-path problems~\cite{Melnikova2000,LiboisDavis2022}.
The slab problem is traditionally approached either by Monte Carlo simulation (MC) or by deterministic radiative-transfer methods (RT)~\cite{Chandrasekhar1960,Stamnes1988}.

We examine the slab problem from the viewpoint of first-passage statistics~\cite{Redner2001, ZellerCordery2020}.
A photon entering a plane-parallel scattering slab performs a persistent random walk~\cite{Chandrasekhar1943}.
Between collisions the step lengths are exponentially distributed, while successive directions are correlated according to the HG phase function.
The photon eventually reaches one of the two slab boundaries and escapes, or is absorbed before escape.
Reflection and transmission are first-passage events of the random walk.
Absorption depends on the path length or the scattering order.

Symmetry between the entry and exit statistics follows from reversibility of the walk.
Reflection can be expressed in terms of a half-space first-return law modified by a finite-slab survival factor~\cite{ZellerCordery2026}.
In the conservative limit, the reflectance tail is linked to the universal \(n^{-3/2}\) first-return tail of random walks.

This paper develops RT and MC as complementary descriptions of this process.
Averaging the HG scattering kernel over azimuth reduces the RT problem to a Markov process in the pair \((z,\mu)\), where \(z\) denotes depth within the slab and \(\mu\) is the direction cosine with respect to the \(z\)-axis.
Confining this process to a finite \(z\) interval with absorbing boundaries produces a transfer operator whose iterates generate the probabilities of escape after a specified number of scattering events.
The MC description is based on generating large ensembles of object excursions using a small number of long random walks.
An ensemble of excursions are generated for several values of \(g\), producing a scattering-order resolved database from which first-passage statistics, angular crossing distributions, and other excursion properties can be extracted directly.
Reflection, transmission, absorption, and emergent (exit) angular distributions follow directly from these order-resolved escape probabilities.

\section{Model and notation}\label{sec:model}

We consider a plane-parallel slab occupying the interval \(0<z<\tau\) .
All distances are measured in units of the scattering mean free path \(\ell\), which we take to be 1. 
The slab coordinate perpendicular to the boundary is \(z\) and \(\tau\) is the slab thickness in scattering-mean-free-path units.

A photon enters the slab at \(z=0\) with incident direction cosine \(\mu_0=\cos\theta_0\). 
Between collisions the step-length \(s \sim \Exp(1/\ell)\). 
If \(\mu_i\) is the direction cosine with the \(z\)-axis during the \(i\)-th flight, the slab coordinate evolves as 
\begin{equation}
	z_{i+1}=z_i+s_i\mu_i.
	\label{eq:zupdate}
\end{equation}
At each collision the propagation direction is redistributed according to the HG phase function
\begin{equation}
	p(\cos\theta;g) = \frac{1}{2}\,	\frac{1-g^2}{(1+g^2-2g\cos\theta)^{3/2}},
	\qquad
	\int_{-1}^{1}p(\cos\theta;g)\,\dif\cos\theta=1 .
	\label{eq:phg}
\end{equation}
The parameter \(g=\avg{\cos\theta}\) is the scattering asymmetry. 
Preferential forward scattering increases the transport mean free path to 
\begin{equation}
	\lstar=\frac{\ell}{1-g},
	\label{eq:lstar}
\end{equation}
the characteristic distance over which the initial propagation direction is effectively forgotten.

We use \(R,T\), and \(A\) for reflectance, transmittance, and absorptance. 
The probability that a photon survives a collision rather than being absorbed is \(a\). 
The scattering order \(n\) is the number of collisions before escape. 
The direction cosine is denoted by \(\mu\), and \(\mu_0\) is reserved for the incidence cosine.

The segment of the walk between a crossing of a slab surface and its first exit from the slab is called an excursion. 
It is transmitted if it reaches \(z=\tau\) and reflected if it returns to \(z=0\). 
The corresponding probabilities for entry cosine \(\mu\) are denoted by \(t(\mu,\tau)\) and \(r(\mu,\tau)=1-t(\mu,\tau).\)

Absorption is described here by the single-scattering albedo $a$, the probability of a photon surviving a collision. 
In \cite{ZellerCordery2020} we used the absorption rate $\chi$. 
Here we work with the equivalent survival probability $a$. 
Due to the memoryless property of $s\sim\Exp(1/\ell)$, these two approaches are equivalent. 
Collisions occur as a Poisson process of rate $\ell^{-1}$ along the trajectory.
If each collision is independently a scatter with probability $a$ and an absorption with probability $1-a$, the Poisson thinning theorem guarantees that the absorption events themselves form an independent Poisson process, with rate $\chi = (1-a)/\ell$. 

\section{Two methods}\label{sec:methods}

We analyze the slab first passage process using two complementary methods. 
The deterministic RT calculation and the direct MC sampling describe the same HG random walk model in a finite slab. 
The two methods agree within numerical and statistical uncertainties specific to each approach.

\subsection{Radiative-transfer formulation}

The RT approach follows the evolution of the angular intensity in a finite plane-parallel slab.
The escape problem depends only on the sequence $\{z_k\}$ and on the direction cosines that generate it. 
The azimuthal angle also evolves, but it affects only the transverse coordinates and is irrelevant to escape from a plane-parallel slab. 
The Markov state is the pair $(z,\mu)$. 
For $g>0$, the projected coordinate $z$ retains directional memory, whereas the pair $(z,\mu)$ is a closed Markov process.
The RT approach is not dependent on a memoryless step length distribution. \(\Exp\) is used here for comparison with the MC approach. 

Averaging the HG phase function over the azimuth reduces the transport problem to the state variables $(z,\mu)$, where $z$ is the slab coordinate and $\mu$ the direction cosine. 
Reflection and transmission correspond to escape through the boundaries $z=0$ and $z=\tau$.

The transport equation is solved deterministically in the confined slab geometry.
Its solution provides the reflected and transmitted fluxes together with their angular distributions. 
Absorption is introduced through the survival probability $a$, allowing reflection, transmission and absorption to be evaluated using the scattering order expansion of the conservative transport operator.

The detailed construction of the confined $(z,\mu)$ transfer operator, together with the scattering-order decomposition and numerical implementation, is deferred to Section \ref{sec:RTtheory}.

\subsection{Monte Carlo formulation}\label{sec:mc-method}

The Monte Carlo (MC) method generates a number of very long walks in \(\mathbb{R}^3\) as a sequence of straight steps joined at scattering events. 
The walk is independent of any objects such as the slabs considered here. 
The length of each step is sampled from \(\Exp(1)\). 
The polar angle is sampled using the HG phase function Eq. \eqref{eq:phg}. 
The azimuthal angle is uniformly distributed.

Passage into an object occurs where a step in the walk crosses the object boundary in the inward direction. 
Similarly, passage out of an object occurs where the crossing is in the outward direction. 

An excursion is a portion of a long walk inside an object. 
The long walk generates an ensemble of excursions using a collection of overlapping objects in \(\mathbb{R}^3\).  
The MC method depends on the memoryless property of \(\Exp\). 
The portion of a crossing step with one end inside the object has the same distribution as the original walk steps.

At a crossing of a plane with random orientation over the half-sphere, the cosine distribution is the flux-weighted law
\begin{equation}
	p(\mu)=2|\mu|.
	\label{eq:fluxlaw}
\end{equation}

The excursions can be binned or re-weighted to obtain desired angular distributions. 
Below, the objects are slabs with boundaries at integer values of \(z\) with integer thicknesses \(\tau\) up to a chosen maximum. 

\section{The radiative-transfer operator}\label{sec:RTtheory}

The RT method divides the slab into thin layers.
The operator is constructed by composing scattering in the layers. 
The slab of total optical thickness $\tau$ is divided into many thin sublayers of thickness $\Delta\tau$.
The algorithm proceeds layer by layer calculating scattering to first order in $\Delta\tau$. The layers are stacked using a composition rule. 

\subsection{The thin-layer building block}

The Markov chain evolves on the pair $(\mu_i, s_i)$, where $\mu_i$ is the
cosine of the angle between the $i$th propagation direction and the $z$
axis and $s_i$ is the step length. Because the medium is azimuthally
symmetric about $z$, only the polar cosine is dynamically relevant, and
the HG phase function may be integrated over the azimuthal angle $\phi$ about the incoming direction.\cite{EnglerHans2015}

\begin{figure}[ht]\centering
	\includegraphics[width=0.6\textwidth]{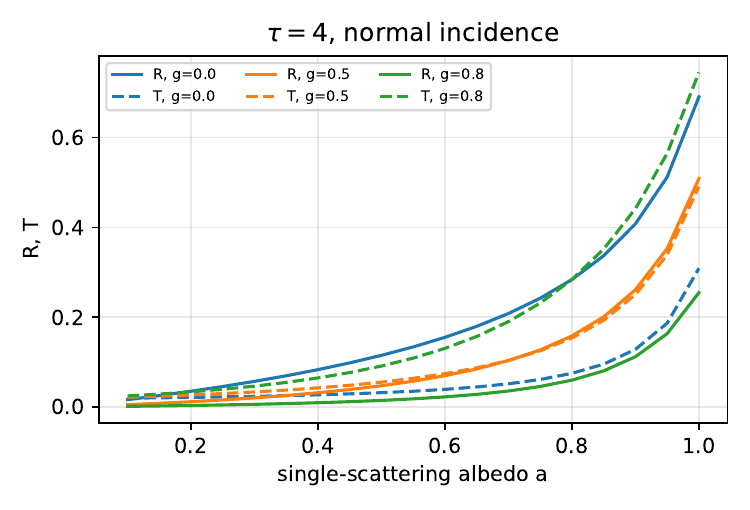}
	\caption{$R,T$ against single-scattering albedo $\tau=4$.}\label{fig:alb}
\end{figure}

\begin{equation}
	p_z(\mu_i,\mu_{i+1}) = \frac{(1-g^2)\,E(k)}{\pi\,(\alpha-\beta)\,\sqrt{\alpha+\beta}},
\end{equation}
where $E(k)$ is the complete elliptic integral of the second kind and
\begin{equation}
	\alpha = 1+g^2-2g\,\mu_i\mu_{i+1},
	\qquad
	\beta = 2g\sqrt{(1-\mu_i^2)(1-\mu_{i+1}^2)},
	\qquad
	k^2 = \frac{2\beta}{\alpha+\beta}.
\end{equation}

For a layer so thin that a photon scatters at most once inside it, the reflection and transmission operators are linear in $\Delta\tau$. Writing the angular redistribution through the Henyey--Greenstein phase function
$p(\mu,\mu')$, a thin layer has
\begin{equation}
	\Top \approx \Id - \frac{\Delta\tau}{\mu}\,
	+ a\,\frac{\Delta\tau}{\mu}\,p_z(\mu,\mu'|\mu\mu'>0), \text{ and}\qquad
	\Rop \approx a\,\frac{\Delta\tau}{\mu}\,p_z(\mu,\mu'|\mu\mu'<0),
\end{equation}
The transmission is mostly the unscattered beam attenuated by $e^{-\Delta\tau/\mu}$, plus the forward-scattered part; 
the reflection is the singly back-scattered fraction, proportional to the single-scattering
survival probability $a$. 
Multiple scattering within the slice is dropped because it is $O(\Delta\tau^2)$.

\subsection{Building the slab}

The slab is built by combining each thin layer with the partial slab already built, using the interaction (adding) equations that re-sum the infinite back-and-forth bouncing between the two pieces:
\begin{eqnarray}
		\Top_{12} =& \Top_2\,\bigl(\Id - \Rop_1\Rop_2\bigr)^{-1}\,\Top_1, \\[4pt]
		\Rop_{12} =& \Rop_1 + \Top_1\,\Rop_2\,\bigl(\Id - \Rop_1\Rop_2\bigr)^{-1}\,\Top_1.
\end{eqnarray}

The factor $(\Id - \Rop_1\Rop_2)^{-1}$ is the geometric series of inter-layer
reflections. The multiple scattering re-enters even though each slice
was treated as 0 or 1 scatter. 
Composing this from $\tau = 0$ up to the full thickness (or \emph{doubling}, when the two stacked pieces are identical) advances the RT function and $\Top$ directly from the RT equation.

\subsection{Absorption}\label{sec:survival}

\begin{figure}[ht]\centering
	\includegraphics[width=0.95\textwidth]{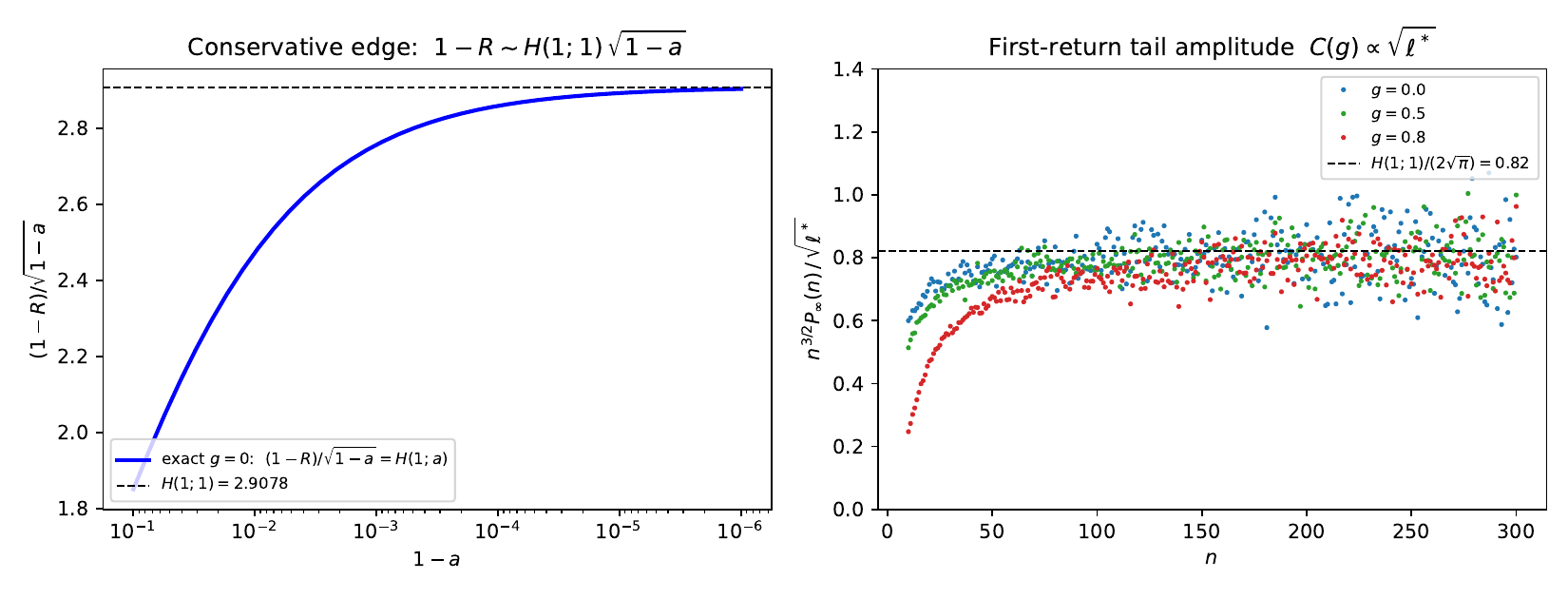}
	\caption{The conservative edge and the first-return tail. \\
		Left: the exact $g=0$ ratio $(1-R)/\sqrt{1-a}=H(1;a)\to H(1;1)=2.9078$ as $a\to1$. \\Right:
		$n^{3/2}\Pinf(n)/\sqrt{\lstar}$ for $g=0,0.5,0.8$ collapses onto
		$H(1;1)/(2\sqrt\pi)=0.82$.}\label{fig:tauber}
\end{figure}

With the single-scattering survival probability $a$, every trajectory of order $n$ survives with probability $a^n$. 
The reflected and transmitted fluxes therefore become
\begin{equation}
	R(a, \tau) = \sum_{n\ge1} P_R(n,\tau)a^n,
\text{ and}\qquad
T(a, \tau) = \sum_{n\ge0}
P_T(n,\tau)a^n .
\label{eq:RT}
\end{equation}

Absorption modifies the trajectory weights of the conservative walk. 
The random walk \((a=1)\) escape probabilities remain unchanged (see Fig.~\ref{fig:alb}).
\begin{equation}
	A(a, \tau) = 1-R(a, \tau)-T(a, \tau)=\sum_{n} (P_R(n,\tau)+P_R(n,\tau))(1-a^n).
	\label{eq:RTAbs}
\end{equation}

\begin{figure}[ht]\centering
	\includegraphics[width=0.6\textwidth]{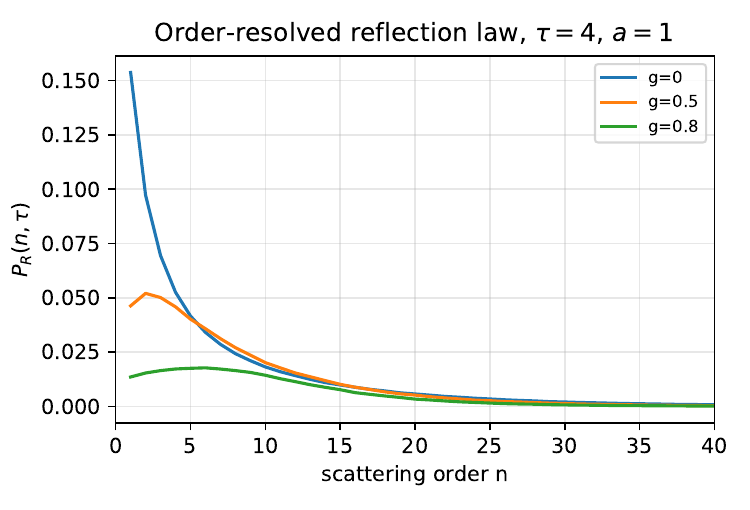}
	\caption{Order-resolved reflection law $\PR(n,\tau)$ at $\tau=4$, $a=1$.}\label{fig:order}
\end{figure}

The order-resolved reflection probability (see Fig.~\ref{fig:order}) admits the decomposition
\begin{equation}
	P_R(n,\tau) = P_\infty(n)\, S(n,\tau),
\end{equation}
where $P_\infty(n)$ is the first-return probability in the semi-infinite
medium and
\begin{equation}
	S(n,\tau)= \Pr(z_{\max}<\tau\,|\,n)
\end{equation}
is the probability that an $n$-step trajectory remains below the upper
boundary.

The corresponding reflectance is therefore
\begin{equation}
	R(a,\tau) = \sum_{n\ge1} P_\infty(n)\, S(n,\tau)\, a^n.
	\label{eq:factor}
\end{equation}
This factorization separates the contributions of the half-space return statistics, the finite-slab geometry, and absorption. The depth-survival factor $S(n,\tau)$ is analyzed in the following section.

All of the thickness dependence in Eq.~\eqref{eq:factor} lives in $S(n,\tau)$.
The depth advances as $z_{k+1}=z_k+s_k\mu_k$ with HG memory
$\mathbb{E}[\mu_{i+k}\mid\mu_i]=g^k\mu_i$, so the free-walk per-step depth variance is
\begin{equation}
	\sigma_g^2=\mathbb{E}[\mu^2]\,\mathbb{E}[s^2]+2\!\sum_{k\ge1}\mathbb{E}[\mu_i\mu_{i+k}]
	=\tfrac23+\tfrac23\frac{g}{1-g}=\tfrac23\,\lstar. 
	\label{eq:sigma}
\end{equation}
Steps are independent, so  $\mathbb{E}[\mu^2]=\tfrac13$, $\mathbb{E}[s]^2=1$ and $\mathbb{E}[s^2]=2$.
In the diffusive regime the conditioned depth process converges to a Brownian excursion, so $S(n,\tau)$ approaches the excursion-maximum law of Kennedy and Chung~\cite{Kennedy1976,Chung1976}, scaling through the combination $\tau/(\sigma_g\sqrt n)$.

\begin{figure}[ht]\centering
	\includegraphics[width=0.9\textwidth]{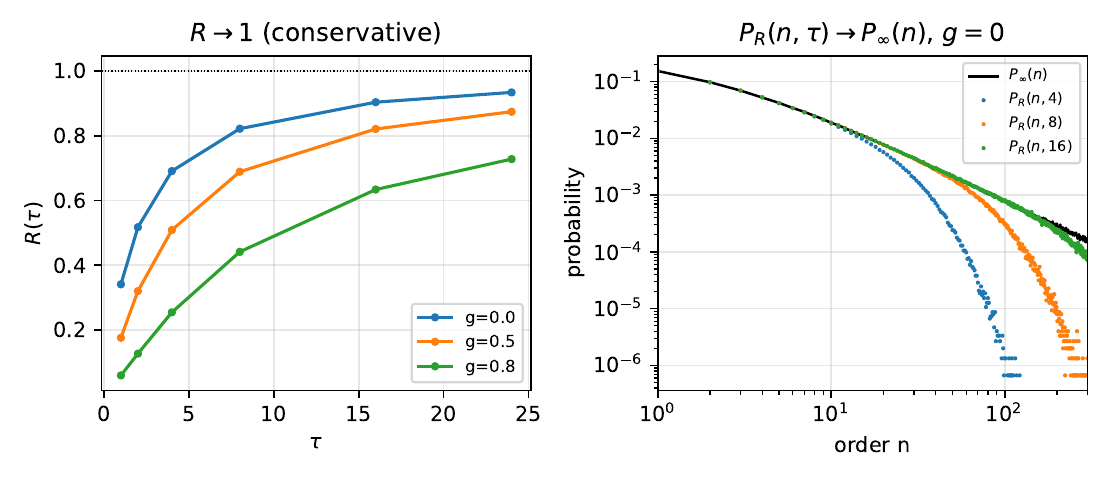}
	\caption{Left: $R(\tau)\to1$ (conservative). Right: $\PR(n,\tau)\to\Pinf(n)$
		($g=0$), peeling off at high order.}\label{fig:thick}
\end{figure}

The half-space reflectance $R(a)=\sum_{n\ge1}\Pinf(n)\,a^n$ rises to its
conservative value $R(1)=1$ as $a\to1^-$, and the manner of approach is fixed by
the large-$n$ tail of the first-return law. That tail carries the universal
Sparre--Andersen exponent \cite{SparreAndersen1953},
\begin{equation}
	\Pinf(n)\sim C(g)\,n^{-3/2}.
	\label{eq:returntail}
\end{equation}

Write the conservative deficit as $1-R(a)=\sum_{n\ge1}\Pinf(n)\,(1-a^n)$, which is
dominated by large $n$ as $a\to1$. Setting $a=e^{-\varepsilon}$ with
$\varepsilon\simeq1-a$ and rescaling $u=n\varepsilon$ turns the sum into
\[
\int_0^\infty u^{-3/2}\bigl(1-e^{-u}\bigr)\,\dif u = 2\sqrt\pi ,
\]
so that
\begin{equation}
	1-R(a)\;\xrightarrow{\,a\to1^-\,}\;2\sqrt\pi\,C(g)\,\sqrt{1-a}.
	\label{eq:tauber}
\end{equation}
The reflectance therefore reaches unity through a square-root \emph{edge}: 
a non-analytic $\sqrt{1-a}$ cusp of infinite slope at $a = 1$, with coefficient $K(g)=2\sqrt\pi\,C(g)$. 
The edge exponent $\tfrac12$ is universal, fixed by the divergence of the mean number of collision before first return, driven by the exponent $\tfrac32$ in Eq.~\eqref{eq:returntail}.

The isotropic case is exact and determines the constants. 
At $g=0$, $R(a)=1-\sqrt{1-a}\,H(1;a)$ with $H$ the conservative Chandrasekhar $H$-function, giving $K(0)=H(1;1)=2.9078$ and $C(0)=H(1;1)/(2\sqrt\pi)=0.820$. 
Our operator tail gives $C(0)=0.79$. 
For $g>0$ the amplitude follows the transport scaling $C(g)\propto\sqrt{\lstar}=(1-g)^{-1/2}$.
The measured $C(g=0.5)/C(0)=1.40\,\approx\sqrt{\lstar}=1.41$. The correction to the asymptotic tail is of relative order \(\lstar/n\), slower at \(g = 0.8\) by a factor 2.5, so $C(g=0.8)/C(0)=2.13\,<\sqrt{\lstar}=\,2.24$ (see Fig.~\ref{fig:tauber}).

\subsection{Thick-slab limit: recovery of the half-space return law}\label{sec:thick}
\begin{figure}[ht]\centering
	\includegraphics[width=0.6\textwidth]{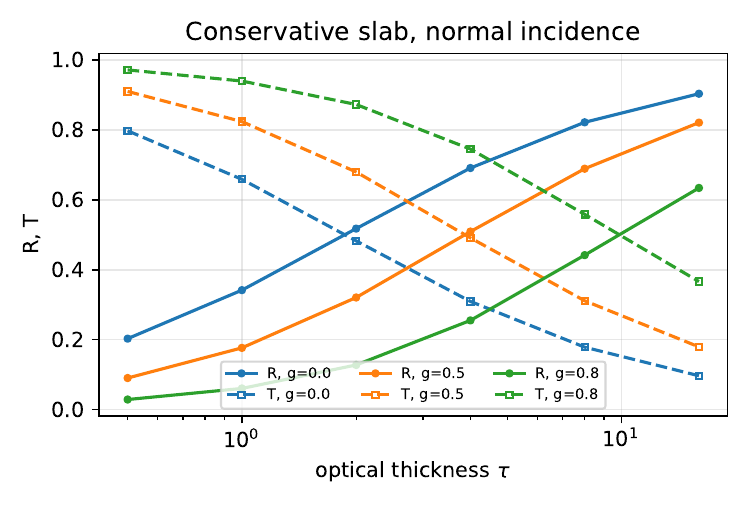}
	\caption{$R,T$ against optical thickness $\tau$ ($a=1$, normal incidence).}\label{fig:tau}
\end{figure}

As $\tau\to\infty$ the conservative slab returns every photon, so as $R\to1$, according to the heavy-tailed behavior of a random walk (gambler's ruin), $T\sim 1/\tau$ for thickness much larger than $\lstar$ (Table~\ref{tab:cons}; see Fig.~\ref{fig:tau}).
The order-resolved statement follows from the factorization (\ref{eq:factor}): the slab law $\PR(n,\tau)$ coincides with $\Pinf(n)$ wherever the back wall is invisible ($S\to1$), peeling away only beyond an order
$n^\ast(\tau)$ that marches outward as $\tau$ grows.
Numerically the thick slab recovers $\Pinf(n)$ term by term, including its $n^{-3/2}$ tail (see Fig.~\ref{fig:thick}).
This order-resolved recovery is the sum rule (\ref{eq:sumrule}) resolved by scattering
order: 
as the back wall recedes the transmission channel closes ($q\to0$) and the full $2\mu$ return flux is rebuilt.

\begin{figure}[ht]\centering
	\includegraphics[width=0.6\textwidth]{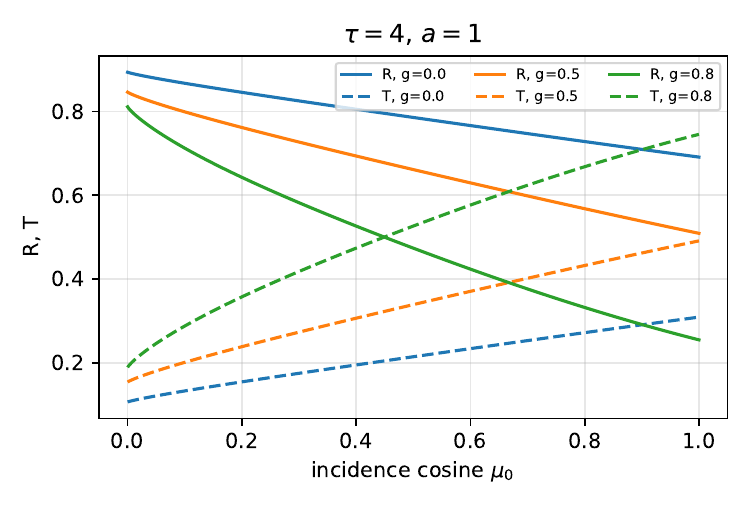}
	\caption{$R,T$ against incidence cosine $\mu_0$ at $\tau=4$, $a=1$.}\label{fig:ang}
\end{figure}

\begin{table}[h]\centering
	\begin{tabular}{ccccccc}
		\toprule
		& \multicolumn{2}{c}{$g=0$} & \multicolumn{2}{c}{$g=0.5$} & \multicolumn{2}{c}{$g=0.8$}\\
		\cmidrule(lr){2-3}\cmidrule(lr){4-5}\cmidrule(lr){6-7}
		$\tau$ & $R$ & $T$ & $R$ & $T$ & $R$ & $T$\\
		\midrule
		1  & 0.3413 & 0.6587 & 0.1761 & 0.8239 & 0.0600 & 0.9400\\
		2  & 0.5175 & 0.4825 & 0.3203 & 0.6797 & 0.1272 & 0.8728\\
		4  & 0.6909 & 0.3091 & 0.5090 & 0.4910 & 0.2547 & 0.7453\\
		8  & 0.8218 & 0.1782 & 0.6890 & 0.3110 & 0.4416 & 0.5584\\
		16 & 0.9036 & 0.0964 & 0.8210 & 0.1790 & 0.6339 & 0.3661\\
		32 & 0.9497 & 0.0502 & 0.9031 & 0.0968 & 0.7836 & 0.2164\\
		\bottomrule
	\end{tabular}
	\caption{Conservative slab reflectance/transmittance against optical thickness (Eq.~\ref{eq:TR_diff})
		($a=1$, normal incidence using RT method, $|R+T-1|\le5\times10^{-5}$).
		For $\tau>2\lstar, \,T\approx1.68/(\tau(1-g)+2z_0)$ where \(z_0\approx0.71\ell\) is the Milne extrapolation length~\cite{Chandrasekhar1960}} and
	\label{tab:cons}
\end{table}

\section{Monte Carlo random walk method}

Our method uses a Monte Carlo random walk to estimate the statistics of excursions on objects in 3D space.
We generate a small number of long random walks with \(s_i\sim\Exp(1)\) and orientations generated by HG.
The random walk is independent of the objects geometry. 

For a slab of thickness $\tau$, first passage occurs where a segment first returns to the initial surface at $z=z_i$ and is a reflection event or crosses $z=z_i+\tau$ and is a transmission event. 
For a half space, although walks are guaranteed to eventually return to the initial surface, the average number of steps is infinite. For a large object the excursion may terminate inside the object due to a limit imposed on the number of steps. 
The scattering order $n$,  the entrance and exit cosines, and path length of the excursion are recorded.
The same ensemble therefore gives reflection and transmission probabilities, angular crossing laws, scattering order distributions, and excursion statistics. 

Absorption is introduced to the conservative ensemble by assigning the survival weight $a^n$ to an excursion of order-$n$.
This is the Monte Carlo counterpart of Eq. \eqref{eq:RT}, keeping the first-passage geometry separate from absorption.

After the walk length is much longer than \(\lstar\), the step orientation is randomized over the sphere.
A step is more likely to intersect a surface if it is a long step with direction close to the boundary normal. 
As the walk crosses a boundary into the object, its direction cosine \(\mu\) relative to the surface normal is distributed by \(p(\mu)=2|\mu|\). 
If the application requires a different distribution of \(\mu\), the direction cosines are binned and re-weighted. 

After detecting crossings of the walk with surfaces of the objects, sequential crossings on the same object are used to construct excursions. The necessary details of each excursion is saved in an array, including initial and final angles, step numbers, path length and initial and final position in \(\mathbb{R}^3\). 

\begin{figure}[ht]\centering
	\includegraphics[width=0.9\textwidth]{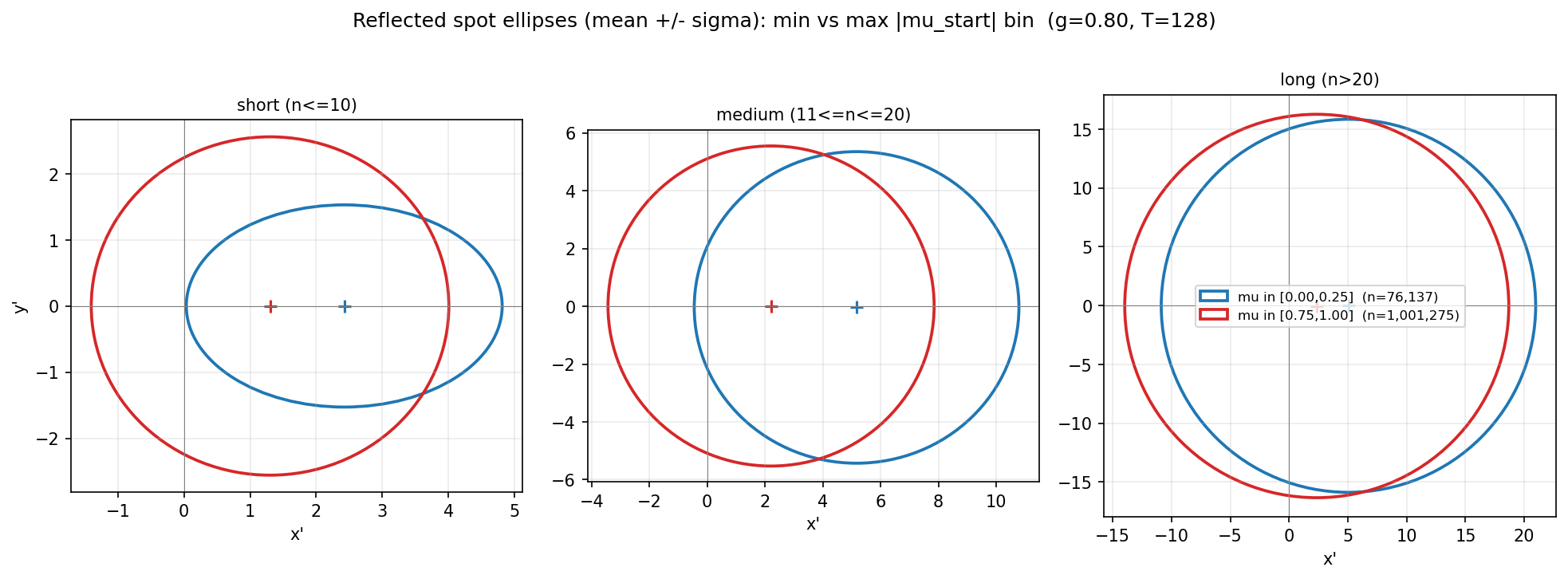}
	\caption{g=0.80 Spot position and displacement in reflection for a thick slab with short, medium and long excursions. Red is \(\mu_{in}>0.75\) and blue is \(\mu_{in}<0.25 \).}\label{fig:ellipses_reflected_g0_80}
\end{figure}

The recorded details give statistics of interest, including, for example, shift in the exit point with initial angle (see Fig.~\ref{fig:ellipses_reflected_g0_80}). 

\begin{figure}[ht]\centering
	\includegraphics[width=0.9\textwidth]{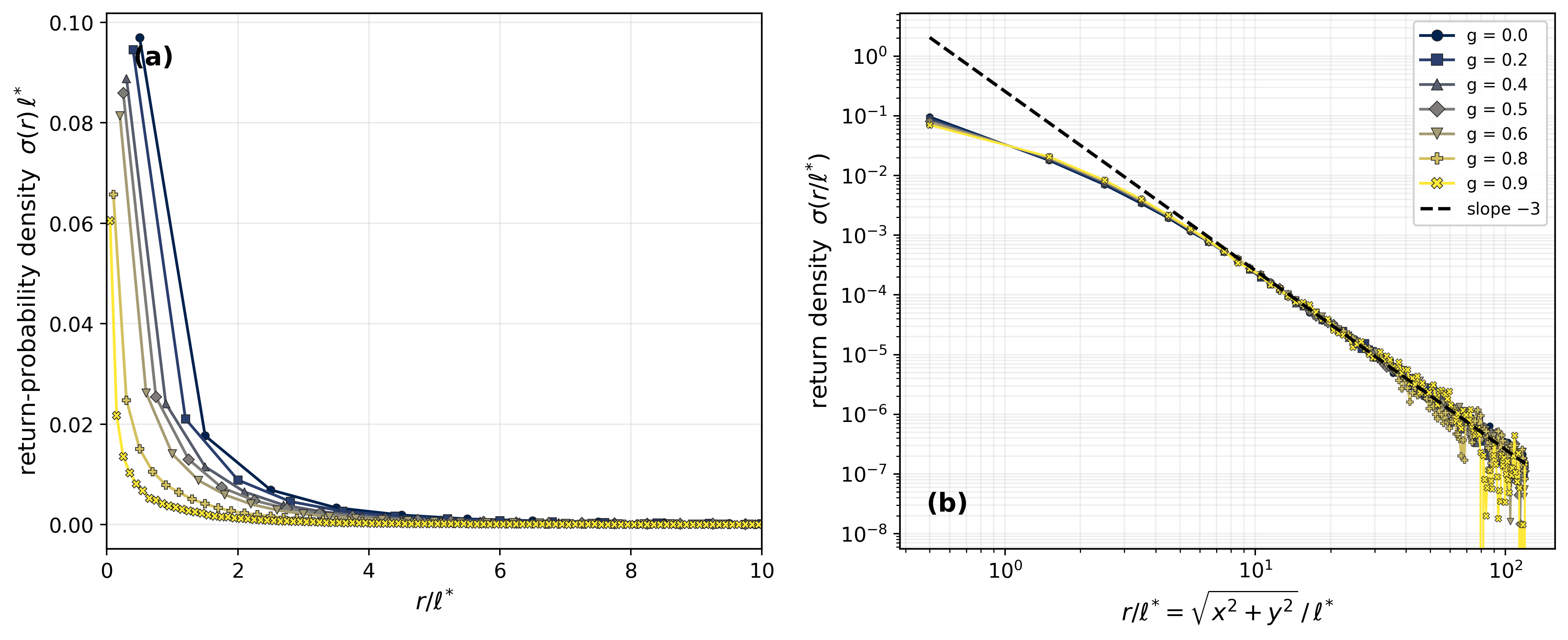}
	\caption{g=0.80 Point-spread-function in reflection showing the collapse to the power law for radius much larger than \(\lstar\).}\label{fig:psf_collapse}
\end{figure}

A second example is the point-spread-function in more detail with the dependence on \(g\). The radius scaled by \(\lstar\) collapses the distribution tail for all values of \(g\) when \(r>\lstar\) (see Fig.~\ref{fig:psf_collapse}).

Using quaternions for rotation of directions is efficient and numerically stable.
The walker state at step $k$ is \(\left(\bm{r}_k,\ q_k\right)\) where \(\bm{r}_k\) is the position and \(q_k\) is a unit quaternion rotating the lab frame to the step direction frame so that the $\hat{z}$ axis is the current direction of travel. 
The lab direction is recovered by conjugation, 
\(\bm{n}_k = q_k\,\hat{z}\,q_k^{*}\)\label{eq:dir}.
Each collision is a rotation in the photon's own frame using the HG distribution by right multiplication
\begin{equation}
	q_{k+1}=q_k \otimes \left(\cos\tfrac{\theta}{2},\
	\sin\tfrac{\theta}{2}\,(-\sin\varphi,\ \cos\varphi,\ 0)\right) .
	\label{eq:compose}
\end{equation}
The deflection is sampled relative to the current heading, and \eqref{eq:compose} folds it into the slab-frame without forming a rotation matrix.

\section{Time reversal, reciprocity, and the sum rule}\label{sec:symmetry}

Because the Henyey--Greenstein phase function depends only on the scattering angle, the walk is reversible.
Combined with the mirror symmetry of the slab, this gives symmetric weights for \( (\mu_i,\mu_f)\) and \((\mu_f,\mu_i)\) both in reflection and transmission. 

The quantities 
\begin{equation}
	t(\mu,\tau) = P(\text{first exit through }z=\tau\mid\mu_0=\mu),
\end{equation}
and
\begin{equation}
	r(\mu,\tau) = P(\text{first exit through }z=0\mid\mu_0=\mu)
\end{equation}
are the conditional first-passage probabilities for a photon entering the slab with direction cosine $\mu$. 
Since every photon must eventually leave through one of the two faces, 
\begin{equation}
	t(\mu,\tau)+r(\mu,\tau)=1.
	\label{eq:sumrule}
\end{equation}

The angular distribution in a fully developed random walk is uniform over the sphere.  The direction cosine distribution for a step crossing the surface is then \(p(\mu)=2|\mu|\).
This equilibrium crossing law is independent of slab thickness and of the Henyey--Greenstein asymmetry parameter.
The transmitted and reflected exit-angle distributions are therefore 
\begin{equation}
	p_{\rm tran}(\mu)
	=
	\frac{2\mu\,t(\mu,\tau)}{T_{\rm diff}(\tau)}, \text{ and } \qquad
	p_{\rm refl}(\mu)
	=
	\frac{2\mu\,r(\mu,\tau)}{R_{\rm diff}(\tau)},
\end{equation}
where
\begin{equation}
	T_{\rm diff}(\tau)
	=
	\int_0^1
	2\mu\,t(\mu,\tau)\,d\mu,\text{ and } \qquad 
	R_{\rm diff}(\tau)
	=
	\int_0^1
	2\mu\,r(\mu,\tau)\,d\mu.
	\label{eq:TR_diff}
\end{equation}
The dependence on slab thickness and anisotropy appears only in this partition.

\section{Validation}\label{sec:validation}

\begin{figure}[ht]\centering
	\includegraphics[width=0.9\textwidth]{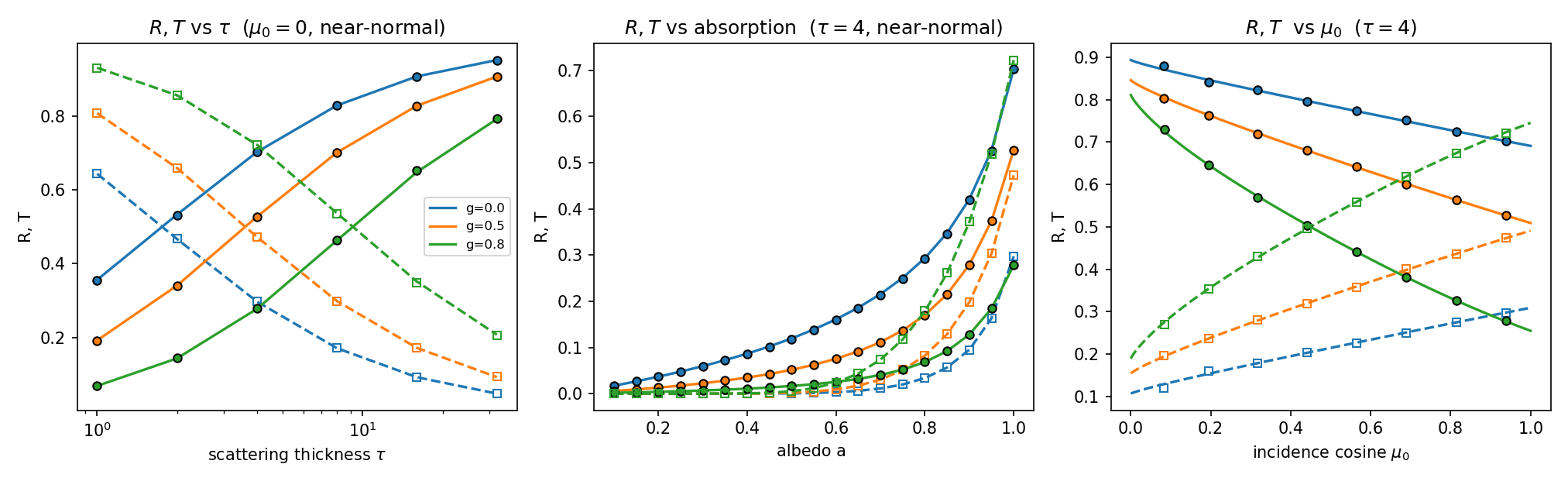}
	\caption{RT and MC comparison. (\(R\) solid, \(T\) dashed)}\label{fig:rw3dfit}
\end{figure}

The forward operator reproduces a full three-dimensional HG random-walk Monte
Carlo in both channels (see Fig.~\ref{fig:rw3dfit}) to $\le1.6\times10^{-3}$ (absolute) over twelve cases
spanning $g\in[0,0.95]$, $\tau\in[0.5,16]$, $a\in[0.5,1]$, and normal-to-oblique
incidence (see Fig.~\ref{fig:ang}), with conservation $R+T=1$ recovered to within $10^{-4}$ at $a=1$; the
Monte-Carlo seed scatter ($\le1.4\times10^{-3}$) is itself at this level
(table \ref{tab:mc}).
The operator is also consistent with two independent
deterministic slab solvers---an adding--doubling solver, which builds a thick slab
by repeatedly stacking and combining identical thin layers
\cite{vandeHulst1980,deHaan1987}, and a successive-orders solver, which sums the
contributions of each scattering order from an integral equation
\cite{vandeHulst1980}---together with the conservative Chandrasekhar $H$-function.
These three references agree among themselves to $\le1.1\times10^{-4}$; because
they share no algorithm with the random walk, their agreement rules out any bias
common to the Monte Carlo, in the spirit of the standard RT
benchmarks \cite{GarciaSiewert1985}.
For the symmetry results of section
\ref{sec:symmetry}, an isotropic $g=0$ Monte Carlo with $4\times10^6$ excursions
confirms reciprocity (joint-kernel asymmetry under $1\%$) and the sum rule (total
exit density flat to within $0.3\%$ at $\tau=1$ and $5$), even while the
individual channels remain strongly biased in $\mu$.

\begin{table}[h]\centering
\begin{tabular}{cccccccc}
\toprule
$g$ & $\tau$ & $a$ & $\theta_0$ & $R_{\rm op}$ & $R_{\rm MC}$ & $|\Delta R|$ & $|\Delta T|$\\
\midrule
0    & 4 & 1   & 0  & 0.6908 & 0.6892 & 0.0016 & 0.0016\\
0.5  & 4 & 1   & 0  & 0.5090 & 0.5085 & 0.0004 & 0.0004\\
0.8  & 4 & 1   & 0  & 0.2548 & 0.2545 & 0.0003 & 0.0003\\
0.95 & 4 & 1   & 0  & 0.0544 & 0.0545 & 0.0001 & 0.0001\\
0.5  & 4 & 1   & 60 & 0.6610 & 0.6608 & 0.0003 & 0.0003\\
0.5  & 16& 1   & 0  & 0.8210 & 0.8203 & 0.0007 & 0.0007\\
0.5  & 4 & 0.9 & 0  & 0.2611 & 0.2606 & 0.0005 & 0.0001\\
0.8  & 8 & 0.7 & 0  & 0.0365 & 0.0363 & 0.0002 & 0.0002\\
\bottomrule
\end{tabular}
\caption{Operator against three-dimensional Monte Carlo (representative subset;
$N=4\times10^5$ photons, $\theta_0$ in degrees).}
\label{tab:mc}
\end{table}

\begin{figure}[ht]\centering
	\includegraphics[width=0.95\textwidth]{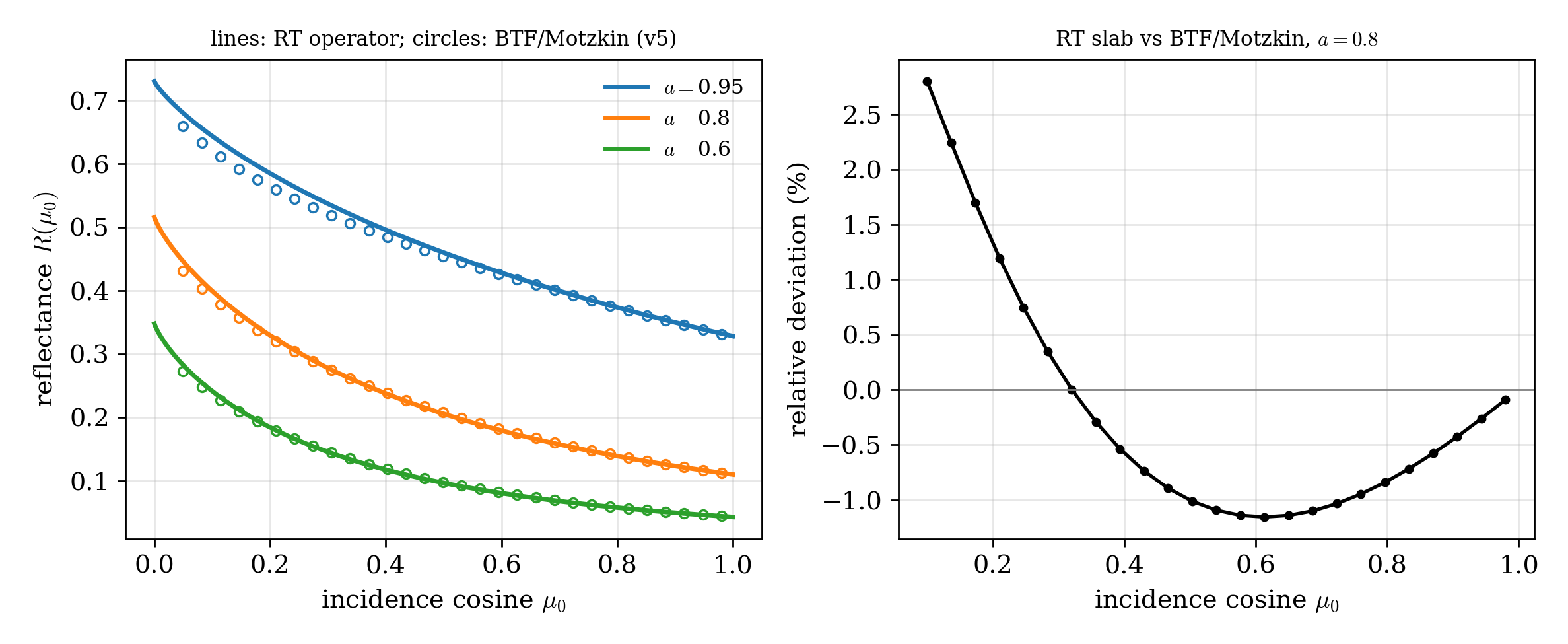}
	\caption{Left: thick-slab angular reflectance $R(\mu_0)$ at $g=2/3$ from the RT
	operator (lines) against the semi-infinite BTF/Motzkin formula of
	\cite{ZellerCordery2026} (circles), for $a=0.95,0.8,0.6$. Right: relative deviation at $a=0.8$.}\label{fig:v5check}
\end{figure}

As a further, independent check we compared the thick-slab ($\tau=64$, effectively
semi-infinite) angular reflectance $R(\mu_0,a)$ from the RT operator against the
closed-form oblique-incidence Boundary-Truncation-Factor/Motzkin formula of
\cite{ZellerCordery2026}, whose oblique extension had not previously been checked
against an independent method away from normal incidence. After reconciling a
scattering-order offset between the two conventions ($n$ here equals their $n-1$),
the two frameworks agree to $\lesssim2\%$ over most of the $(\mu_0,a)$ range,
degrading to $3$--$10\%$ only near grazing incidence ($\mu_0\lesssim0.3$) and high
albedo ($a\gtrsim0.9$), Fig.~\ref{fig:v5check}, consistent with the Cauchy-kernel
fit tolerance quoted there and closing the oblique-incidence validation gap left
open in that paper.

\section{Discussion}\label{sec:disc}

A single depth walk underlies the whole calculation, and the connecting identities
are explicit. The stationary walk is reversible, which gives the reciprocity
between entry and exit statistics; its equilibrium directions are isotropic, which
fixes $2|\mu|$ as the crossing law (Sec.~\ref{sec:symmetry}). The same walk,
confined to $0<z<\tau$ and killed at the faces, is the transfer operator whose
order-resolved escape probabilities give $R$ and $T$ (Sec.~\ref{sec:RTtheory}).
Absorption enters as the per-collision weight $a^n$, and its conservative
$\sqrt{1-a}$ edge is set by the $n^{-3/2}$ first-return tail.
As the slab thickens the transmission channel closes,
$t(\mu,\tau)\to0$, and each order's reflection law relaxes to the half-space value,
$\PR(n,\tau)\to\Pinf(n)$ (Sec.~\ref{sec:thick}).

Several directions remain open.
\begin{itemize}
	\item 	The full angular output---the bidirectional reflectance and transmittance
	distributions---from the higher azimuthal modes $m>0$ of the redistribution
	kernel, which decouple into independent operators of the same form as the $m=0$
	case treated here.
	\item A closed form for the leading eigenvalue $\lambda_1(\tau,g)$ of the confined
	operator, which fixes the asymptotic decay of the intensity with scattering order.
	\item Absorption carried inside the diffusion law of Sec.~\ref{sec:survival}, where
	a screened equation $\phi''=\kappa^2\phi$ replaces the algebraic $1/\tau$
	transmission with an exponential decay $e^{-\tau/L_d}$ over a diffusion length
	$L_d=1/\kappa$.
	\item The strongly forward-peaked regime $g\to1$, relevant to tissue, snow, and
	cloud, which needs finer $\mu$ resolution or a $\delta$-$M$ truncation of the
	forward peak.
\end{itemize}


\begin{thebibliography}{99}

\bibitem{HenyeyGreenstein1941}
Henyey L G and Greenstein J L 1941 Diffuse radiation in the galaxy
{\em Astrophys.\ J.} {\bf 93} 70--83

\bibitem{Chandrasekhar1960}
Chandrasekhar S 1960 {\em Radiative Transfer} (New York: Dover)

\bibitem{Chandrasekhar1943}
Chandrasekhar S 1943 Stochastic problems in physics and astronomy
{\em Rev.\ Mod.\ Phys.} {\bf 15} 1--89

\bibitem{vandeHulst1980}
van de Hulst H C 1980 {\em Multiple Light Scattering: Tables, Formulas and
Applications} vol 1--2 (New York: Academic)

\bibitem{Stamnes1988}
Stamnes K, Tsay S-C, Wiscombe W and Jayaweera K 1988 Numerically stable
algorithm for discrete-ordinate-method radiative transfer in multiple scattering
and emitting layered media {\em Appl.\ Opt.} {\bf 27} 2502--9

\bibitem{Melnikova2000}
Melnikova I N, Dlugach Zh M, Nakajima T and Kawamoto K 2000 Calculation of the
reflection function of an optically thick scattering layer for a
Henyey--Greenstein phase function {\em Appl.\ Opt.} {\bf 39} 4195--204

\bibitem{LiboisDavis2022}
Libois Q and Davis A B 2022 Photon path distributions in optically thin slabs
{\em Opt.\ Express} {\bf 30} 40968--90

\bibitem{SparreAndersen1953}
Sparre Andersen E 1953 On the fluctuations of sums of random variables
{\em Math.\ Scand.} {\bf 1} 263--85; 1954 {\em Math.\ Scand.} {\bf 2} 195--223

\bibitem{Redner2001}
Redner S 2001 {\em A Guide to First-Passage Processes} (Cambridge: Cambridge
University Press)

\bibitem{ZellerCordery2020}
Zeller C and Cordery R 2020 Light scattering as a Poisson process and
first-passage probability {\em J.\ Stat.\ Mech.} {\bf 2020} 063404

\bibitem{ZellerCordery2026}
Zeller C and Cordery R 2026 First-return statistics in Henyey--Greenstein
scattering: Motzkin polynomials and the Cauchy kernel {\em J.\ Stat.\ Mech.}
{\bf 2026} 043206

\bibitem{Kennedy1976}
Kennedy D P 1976 The distribution of the maximum Brownian excursion
{\em J.\ Appl.\ Probab.} {\bf 13} 371--6

\bibitem{Chung1976}
Chung K L 1976 Excursions in Brownian motion {\em Ark.\ Mat.} {\bf 14} 155--77

\bibitem{deHaan1987}
de Haan J F, Bosma P B and Hovenier J W 1987 The adding method for multiple
scattering calculations of polarized light {\em Astron.\ Astrophys.} {\bf 183}
371--91

\bibitem{GarciaSiewert1985}
Garcia R D M and Siewert C E 1985 Benchmark results in radiative transfer
{\em Transp.\ Theory Stat.\ Phys.} {\bf 14} 437--83

\bibitem{EnglerHans2015}
Engler H 2015 Computation of scattering kernels in radiative transfer
{\em J.\ Quant.\ Spectrosc.\ Radiat.\ Transfer} {\bf 165} 38--42

\end{thebibliography}
\end{document}